\documentclass[letter]{jpsj2} 
\title{Realization of Heavy Local Fermi Liquid and Non-Fermi Liquid in f$^2$ Crystalline-Electric-Field Singlet-Triplet Configuration}

\author{Kazumasa \textsc{HATTORI} and Kazumasa \textsc{MIYAKE}}

\inst{Division of Materials Physics, Department of Materials Engineering Science, Graduate School of Engineering Science, Osaka University, Toyonaka, Osaka 560-8531, Japan}

\abst{Using the numerical renormalization group method, we investigate an extended Anderson model, in which correlated electrons with the $\Gamma_1$(singlet)-$\Gamma_4$(triplet) f$^2$ crystalline-electric-field (CEF) configuration hybridize with conduction electrons of $\Gamma_7$(doublet) and $\Gamma_8$ (quartet) under cubic $O_h$ symmetry, from a strong spin-orbit interaction limit. For the case of the parameters relevant to PrFe$_4$P$_{12}$, the system is under competition between the CEF singlet fixed point and the multichannel Kondo non-Fermi liquid fixed point arising from the quadrupolar coupling between the impurity with pseudospin 1 and the conduction electron with pseudospin 3/2 . We consider that this result reveals the origin of the heaviness of the effective mass and non-Fermi liquid behavior of the Pr-based filled skutterudite compounds.}

\kword{numerical renormalization group, f$^2$ singlet ground state, non-Fermi liquid, conformal field theory, skutterudite, heavy fermion, Kondo effect}

\begin{document}
\maketitle
 Recently, a number of Pr-based filled skutterudite compounds have been investigated both experimentally and theoretically with much attention. Among them, PrOs$_4$Sb$_{12}$\cite{Maple} is attracting much attention because of time-reversal symmetry-breaking superconducting states\cite{AOKI} and multiple phase diagrams\cite{TAYAMA}.  Since the crystalline-electric-field (CEF) level for PrOs$_4$Sb$_{12}$ has been well established \cite{KOHGI,Goremychkin, Goto} by both neutron scattering and ultrasonic experiments, it is important to know what happens under this CEF level scheme, in which the ground state is a $\Gamma_1$ singlet with the lowest excited state is a $\Gamma_4$ triplet. f-electron physics under CEF exhibits many interesting properties such as a two-channel (quadrupolar) Kondo effect\cite{COX,COX-ZAWA} and competition between the CEF and Kondo-Yosida singlet\cite{YOTSU}.

 In this Letter, we study an extended impurity Anderson model with the f$^2$-CEF scheme of $\Gamma_1-\Gamma_4$ hybridizing with conduction electrons of $\Gamma_7$ and $\Gamma_8$ symmetry. Recently, several studies using only a $\Gamma_7$ conduction electron band have been carried out\cite{Shiina,Otsuki}, since it appears according to band calculations\cite{Harima} that the conduction electrons in Pr-based filled skutterudites mainly come from the pnictogen molecular orbital with $A_{1u}$ (i.e., $\Gamma_7$) symmetry. However, except for PrRu$_4$Sb$_{12}$, there are certain amounts of contributions from the transition metal atoms \cite{HarimaPriv}, e.g., PrOs$_4$Sb$_{12}$ and PrFe$_4$P$_{12}$. It is natural to think that each filled skutterudite compound can be distinguished not only by the CEF levels but also by the variety in the conduction electron bands. Then it is not unrealistic to investigate a model including both $\Gamma_7$ and $\Gamma_8$ conduction electrons particularly as an impurity problem. 

 First, let us start by introducing an extended impurity Anderson Hamiltonian appropriate to the present case. We restrict ourselves within the Hilbert space that includes $\Gamma_1$ (singlet) and $\Gamma_4$ (triplet) states for the f$^2$-CEF configurations and $\Gamma_7$ (doublet) and $\Gamma_8$ (quartet) states for f$^1$-CEF ones under $O_h$ symmetry. Hybridizations $v_{\Gamma_7}$ and $v_{\Gamma_8}$ are expressed easily if we work in a j-j coupling scheme with $J=5/2$ manifold. Although there is no mixing between $\Gamma_4$ and $\Gamma_5$ states in $O_h$ symmetry (the finite mixing is a striking feature in $T_h$ symmetry\cite{Th}), we expect that such a obstacle will not hinder us from discussing the qualitative properties of a system with small deviations from $O_h$ symmetry such as PrFe$_4$P$_{12}$. Even in the j-j coupling scheme, which is valid in a strong spin-orbit limit, we can include the effects of Hund's rule coupling in our model by considering the f$^2$ states with the total angular momentum $J=4$. 

 Thus, we state the model Hamiltonian as
\begin{eqnarray}
H= H_{\rm c}+H_{{\rm f}^1}+H_{{\rm f}^2}+H_{{\rm hyb}},\label{IAM}\hspace{2.5cm}
\end{eqnarray}
\begin{eqnarray}
H_{\rm c}=\sum_{k}\Big[\sum_{\nu=\uparrow,\downarrow}\epsilon_k^{\Gamma_7}c^{\dagger}_{k\nu}c_{k\nu}+\!\!\!\!\!\!\sum_{\mu=\pm\frac{3}{2},\pm\frac{1}{2}}\epsilon_k^{\Gamma_8}a^{\dagger}_{k\mu}a_{k\mu}\Big],
\end{eqnarray}
\begin{eqnarray}
H_{{\rm f}^1}=\sum_{i=7,8}\sum_{\rho}\epsilon_{\rm f}^{\Gamma_i}|{\rm f}^1\Gamma_i^{\rho}\rangle \langle {\rm f}^1\Gamma_i^{\rho}|,\hspace{3.5cm}
\end{eqnarray}
\begin{eqnarray}
H_{{\rm f}^2}=E_{\Gamma_1}|{\rm f}^2\Gamma_1\rangle\langle {\rm f}^2\Gamma_1|+\sum_{\alpha=\pm,0}
(E_{\Gamma_1}+\eta)|{\rm f}^2\Gamma_4^{\alpha}\rangle\langle {\rm f}^2\Gamma_4^{\alpha}|,
\end{eqnarray}
\begin{eqnarray}
H_{\rm hyb}\!\!\!\!\!&=&\!\!\!\!\!\sum_{k,\rho}\sum_{i=7,8}\Big\{\sum_{\alpha=\pm,0}\Big[\sum_{\sigma=\uparrow,\downarrow}v_{\Gamma_7\sigma}^{i\rho4\alpha}c_{k\sigma}^{\dagger}
   +\!\!\!\!\!\!\!\!\sum_{\mu=\pm\frac{3}{2},\pm\frac{1}{2}}v_{\Gamma_8\mu}^{i\rho4\alpha}a_{k\mu}^{\dagger}\Big]\nonumber\\
&&\times\ \ |{\rm f}^1\Gamma_i^{\rho}\rangle\langle{\rm f}^2\Gamma_4^{\alpha}|+
\Big[\sum_{\sigma=\uparrow,\downarrow}v_{\Gamma_7\sigma}^{i\rho1}c_{k\sigma}^{\dagger}\nonumber\\
&&+\sum_{\mu=\pm\frac{3}{2},\pm\frac{1}{2}}v_{\Gamma_8\mu}^{i\rho1}a_{k\mu}^{\dagger}\Big]|{\rm f}^1\Gamma_i^{\rho}\rangle\langle{\rm f}^2\Gamma_1|+
{\rm h.c.}\Big\},\label{hyb}
\end{eqnarray}
where the conduction electrons of $\Gamma_8(\Gamma_7)$ are expressed as fermions $a_{k\mu}(c_{k\nu})$ with pseudospin 3/2 (1/2) and the same representation for the f-electron operator $a_{\mu}(c_{\nu})$ is used. $\eta$ is the CEF splitting between the $\Gamma_1$ and $\Gamma_4$ states.
 
We can express two relevant f$^2$ states with the total angular momentum $J=4$ ($\Gamma_1$,\ $\Gamma_4$) as the eigenstates of the total pseudospin $S^{\rm tot}$: $\Gamma_1 \to S^{\rm tot}=0$ and $\Gamma_4 \to S^{\rm tot}=1$ (see Table \ref{tbl-f2}). Then, we can reduce significantly the calculation time and maintain a high accuracy in the numerical renormalization group (NRG) calculation\cite{WILSON}. Namely, we can take advantage of the existence of three conserved quantities: the total pseudospin $S^{\rm tot}$, the total charge $Q^{\rm tot}$ relative to the half filling, and an additional quantity, total {\it helicity} $h^{\rm tot}$, which is a quantity similar to the angular momentum in $C_{\rm 3v}, C_{\rm 4v}$ symmetry\cite{moustakas}, and is defined as
\begin{eqnarray}
h^{\rm tot} = {\rm mod}\{\sum_{\mu}(a_{\mu}^{\dagger}a_{\mu}+\sum_{k}a_{k\mu}^{\dagger}a_{k\mu}),2\}.
\end{eqnarray}
The helicity is introduced to compensate for the lack of charge conservation law for each orbital. It is noted that $h^{\rm tot}$ does not include $c^{\dagger}_{\nu}c_{\nu}$. Then, f$^2$-$\Gamma_1$ states with $J=4$, consisting of the states $c^{\dagger}_{\uparrow}c^{\dagger}_{\downarrow}$ and $(a^{\dagger}_{\frac{1}{2}}a^{\dagger}_{-\frac{1}{2}}-a^{\dagger}_{\frac{3}{2}}a^{\dagger}_{-\frac{3}{2}})$, for example, have a definite helicity $h^{\rm tot}=0$. We can use $h^{\rm tot}$ for sorting states in the NRG calculation, enabling us to diagonalize the block Hamiltonian, at each NRG step, faster by about four times. 

In this Letter, we use the discretization parameter $\Lambda=3$ and perform calculations by retaining up to 1000 states\cite{WILSON}, while we have checked that the results obtained by retaining 2000 states give no difference. For simplicity, we set the density of states of conduction electrons to be constant and the bandwidth $2D$ centered at the Fermi energy for both $\Gamma_7$ and $\Gamma_8$ orbitals. For a later purpose, we define the hybridizations $v_{\Gamma_7}$ and $v_{\Gamma_8}$ as
\begin{eqnarray}
v_{\Gamma_7}\equiv -2v_{\Gamma_7\downarrow}^{8\frac{1}{2}4+},\ \ 
v_{\Gamma_8}\equiv  2v_{\Gamma_8\frac{1}{2}}^{7\uparrow4+}. \label{v}
\end{eqnarray}
Other hybridizations in eq. (\ref{hyb}) are related to $v_{\Gamma_7}$ and $v_{\Gamma_8}$ according to Table \ref{tbl-f2}. Throughout this paper, energy is measured in units of $D$.
\begin{table}[t]
\begin{tabular}{|c|c|c|}
\hline
       & states & $(Q^{\rm tot},S^{\rm tot},S^{\rm tot}_z, h^{\rm tot})$\\
\hline
$\Gamma_7^{\nu}$ & $c^{\dagger}_{\nu}$ & $(-2,\frac{1}{2},\nu,0)$\\
\hline
$\Gamma_8^{\mu}$ & $a^{\dagger}_{\mu}$ & $(-2,\frac{3}{2},\mu,+1)$\\
\hline
$\Gamma_1$ & $\sqrt{\frac{1}{6}}(a^{\dagger}_{\frac{1}{2}}a^{\dagger}_{-\frac{1}{2}}-a^{\dagger}_{\frac{3}{2}}a^{\dagger}_{-\frac{3}{2}})+\sqrt{\frac{2}{3}}c^{\dagger}_{\uparrow}c^{\dagger}_{\downarrow}$ & $(-1,0,0,0)$\\
\hline
$\Gamma_4^{+}$ & $\frac{\sqrt{3}}{2}c_{\downarrow}^{\dagger}a_{\frac{3}{2}}^{\dagger}-\frac{1}{2}c_{\uparrow}^{\dagger}a^{\dagger}_{\frac{1}{2}}$ & $(-1,1,+1,+1)$\\
\hline
$\Gamma_4^{0}$ & $\frac{1}{\sqrt{2}}(c_{\downarrow}^{\dagger}a_{\frac{1}{2}}^{\dagger}-c_{\uparrow}^{\dagger}a^{\dagger}_{-\frac{1}{2}})$ & $(-1,1,0,+1)$\\
\hline
$\Gamma_4^{-}$ & $-\frac{\sqrt{3}}{2}c_{\uparrow}^{\dagger}a_{-\frac{3}{2}}^{\dagger}+\frac{1}{2}c_{\downarrow}^{\dagger}a^{\dagger}_{-\frac{1}{2}}$ & $(-1,1,-1,+1)$\\
\hline
\end{tabular}
\caption{A set of $J=\frac{5}{2}$ (f$^1$) and $J=4$ (f$^2$) states used for present NRG calculations. $\nu=\uparrow,\ \downarrow$ and $\mu=\pm\frac{1}{2},\ \pm\frac{3}{2}$.}
\label{tbl-f2}
\end{table}

Next, to make the physics of this extended Anderson model (EAM) clear, we transform eq. (\ref{IAM}) to a Kondo-like Hamiltonian in the usual way\cite{SWtrans}, assuming $\epsilon_{\rm f}^{\Gamma_j}-E_{\Gamma_i} \gg |v_{\Gamma_{7(8)}}|$ for $i=1,4$ and $j=7,8$.
 The result is 
\begin{eqnarray}
H_{{\rm Kondo}}= [J_{47}{\bf s_7}+J_{48}{\bf s_8}]\cdot {\bf S_4} +J_Q {\bf q_8}\cdot {\bf Q_4}\nonumber\\
+H_{{\rm pot}}+H_{\rm mix}+H_{\rm f^2},\label{Kondo}\\
J_{47}=-\frac{1}{2}\frac{|v_{\Gamma_7}|^2}{\epsilon_{\rm f}^{\Gamma_8}-E_{\Gamma_4}},\ \ \ 
J_{48}=\frac{1}{4}\frac{|v_{\Gamma_8}|^2}{\epsilon_{\rm f}^{\Gamma_7}-E_{\Gamma_4}}\label{Js}
,
\end{eqnarray}
where $2J_Q=J_{48}$. $\bf s_7, s_8$ and $\bf S_4$ represent the pseudospin of the conduction electron with $\Gamma_7, \Gamma_8$ symmetry and that of the impurity with $\Gamma_4$ symmetry, respectively. $\bf q_8$ and $\bf Q_4$ represent the pseudo-quadrupole of the $\Gamma_8$ conduction electron and that of the impurity for $\Gamma_4$ symmetry, respectively. It is noted that ${\bf q_8}$ and ${\bf Q}_4$ are represented by the corresponding pseudospin operators as follows: $ \{S_xS_y+S_yS_x, S_yS_z+S_zS_y, S_zS_x+S_xS_z, S_x^2-S_y^2, (3S_z^2-S^2)/\sqrt{3}\}$. $H_{\rm pot}$ in eq. (\ref{Kondo}) represents the potential scattering at the impurity site. We ignore the k-dependence of all parameters and the small renormalizations of the f$^2$-level $E_{\Gamma_4}$ and $E_{\Gamma_1}$. $H_{\rm mix}$ in eq. (\ref{Kondo}) contains terms such as $\sim a_{\frac{1}{2}}^{\dagger}c_{\downarrow}|\Gamma_1\rangle\langle\Gamma_4^+| $ representing the scattering that mixes $\Gamma_1$ and $\Gamma_4$, and will turn out to play an important role in the stabilization of the CEF singlet. The essential difference from the LS scheme \cite{Shiina, Otsuki} is that it is not possible to rewrite the $H_{\rm mix}$ term using the ``$X$'' operator\cite{Otsuki} in the present model. This is because there is no triplet state constructed by two $\Gamma_7$ electrons in the present scheme. It is noted that $J_{47}<0$ , eq. (\ref{Js}), which means that this coupling is irrelevant as in an usual Kondo model (KM). This point is consistent with the LS scheme.
\begin{figure}
	\begin{center}
    \includegraphics[width=.43\textwidth]{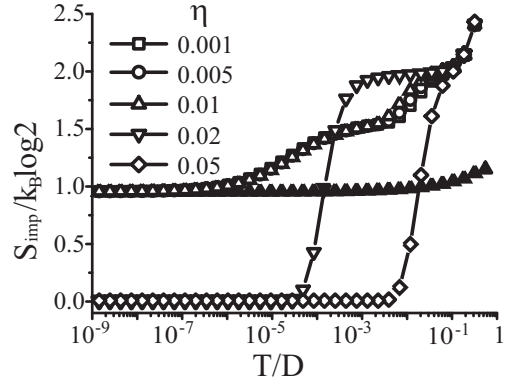}
  \end{center}
\caption{T-dependence of $S_{\rm imp}$, entropy due to an impurity, for EAM (open symbols) and KM (filled triangles). Parameters used for EAM are $E_{\Gamma_{7(8)}}=-0.9$, $E_{\Gamma_1}=-1.5$, $v_{\Gamma_8}=0.3$ and $v_{\Gamma_7}=0.0$. Although the result of EAM is obtained by setting $v_{\Gamma_7}=0$, the qualitative features are the same as in the case of $v_{\Gamma_7}\not= 0$. For KM, we use eq. (\ref{Kondo}) without $H_{{\rm f}^2}$ regarding $J_{48}$ and $J_{Q}$ ($J_{48}=J_{\rm Q}=0.2$) as independent parameters. }
\label{fig-ent}
\end{figure}

 If we ignore $H_{\rm mix}$ and $H_{\rm f^2}$, we can obtain an impurity spin 1 and a conduction electron spin 3/2 multichannel KM with quadrupolar interactions, which was discussed by Koga et al\cite{KOGA}. If we retain only the $J_{48}$ term in eq. (\ref{Kondo}), we can construct an exact boundary conformal field theory (BCFT)\cite{Affleck, Kim,sengupta}, and the non-Fermi liquid (NFL) fixed point spectrum is in complete agreement with the NRG result\cite{KOGA}. If both $J_{48}$ and $J_Q$ terms in eq. (\ref{Kondo}) were considered (no particle-hole symmetry), another NFL fixed point would be obtained, as shown by Koga et al in their NRG calculation. We call this NFL fixed point ``S" following Koga et al. They argued that at the ``S'' fixed point, there exists a pseudo SU(3) symmetry according to the result of the two-loop renormalization group analysis. However, the NRG energy spectrum seems to have no SU(3) symmetry. Detailed analysis of the ``S'' fixed point will be investigated elsewhere\cite{Hat1}. When we include all the terms in eq. (\ref{Kondo}), it is easier to treat the original EAM of eq. (\ref{IAM}) by NRG. Taking the above knowledge into account, it is expected that the interplay between the CEF singlet and the NFL is important in the present model.

\begin{figure}[t]
	\begin{center}
    \includegraphics[width=.43\textwidth]{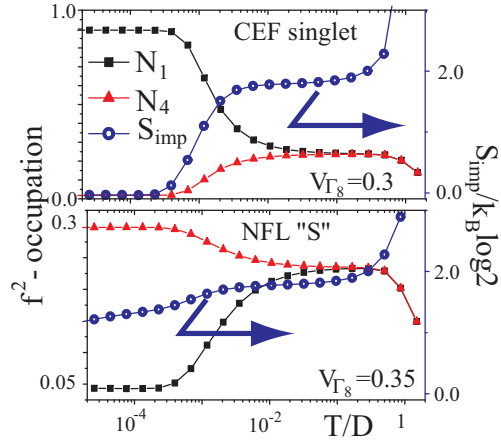}
  \end{center}
\caption{Occupation number of f$^2$-state together with the entropy due to an impurity. The upper panel is for the CEF singlet fixed point ($v_{\Gamma_8}=0.3$) and the lower one is for the S NFL fixed point ($v_{\Gamma_8}=0.35$). Other parameters are $E_{\Gamma_{7(8)}}=-0.2$, $E_{\Gamma_1}=-1.5$, $\eta=0.0005$ and $v_{\Gamma_7}=0.3$. The same symbols are used in both panels. The high-temperature entropy is not $\sim k_B\log4$, because of the truncation procedure. The correct value $S_{\rm imp}={\rm k_B}\log 4$ is recovered when we keep 2000 states in the NRG calculation.}
\label{fig-occu}
\end{figure}

 We show the result of $S_{\rm imp}$, the entropy due to an impurity, in Fig. \ref{fig-ent} together with that of KM without $H_{\rm mix}$ and $H_{{\rm f}^2}$. It is noted that $S_{\rm imp}$ for $\eta=0.001, 0.005, {\rm and\ } 0.01$ in EAM (\ref{IAM}) , and that in KM (\ref{Kondo}) merge into the same value $\sim k_B\log 2$ in the low temperature limit, suggesting that the ground state of EAM with those parameters is the $\Gamma_4$ triplet state. As seen in Fig. \ref{fig-occu}, this is confirmed by the T-dependence of the number of doubly occupied states $N_1$ and $N_4$, which are defined as
\begin{eqnarray}
N_1&\equiv&|{\rm f}^2\Gamma_1\rangle\langle {\rm f}^2\Gamma_1|,\\
N_4&\equiv&\frac{1}{3}\sum_{\alpha=\pm,0}|{\rm f}^2\Gamma_4^{\alpha}\rangle\langle {\rm f}^2\Gamma_4^{\alpha}|.
\end{eqnarray}
 In Fig. \ref{fig-occu}, the T-dependence of $S_{\rm imp}$ is shown together with those of $N_1$ and $N_4$ for typical cases with the CEF singlet and the triplet ground state. A small deviation in $N_1$ and $N_4$  from those ideal values is due to the effect of hybridization with conduction electrons. The meaning of $\lim_{T\to0}S_{\rm imp}\sim k_B\log2$, which is not exactly equal to $k_B\log2$, is unclear for the moment. We can also see the same NRG spectrum in both KM and EAM at the ``S'' fixed point. The NRG spectrum confirms that $k_B\log 2$ does not arise from the contribution of a simple spin $1/2$.

\begin{figure}
	\begin{center}
    \includegraphics[width=.43\textwidth]{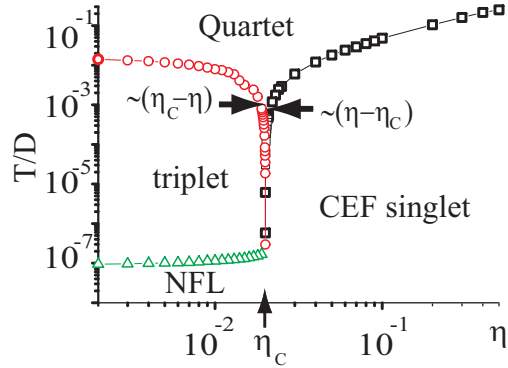}
  \end{center}
\caption{``Phase diagram'' in $T-\eta$ plane for $v_{\Gamma_8}=0.37, v_{\Gamma_7}=0, E_{\Gamma_1}=-1.4 \ {\rm and}\ E_{\Gamma_{7(8)}}=0.2$. The critical value for $\eta$ is determined as $\eta_c=0.02$ for these parameters. At $\eta=\eta_c$, the levels of the f$^2$ singlet and triplet states are interchanged. The crossover temperatures between quartet and triplet states, and between quartet and singlet states are proportional to $|\eta-\eta_c|$ as indicated in the figure.}
\label{fig-PHASE}
\end{figure}

The ``phase diagram'' in the $T-\eta$ plane is shown in Fig. \ref{fig-PHASE}, the phase boundary of which is obtained by estimating the crossover temperature of $S_{\rm imp}$.
 Although we draw a phase diagram only for the parameter set $v_{\Gamma_7}=0$ in Fig. \ref{fig-PHASE}, the qualitative features for $v_{\Gamma_7}\not= 0$ remain the same. The overall features are summarized as follows: a) The NFL ``S'' fixed point is possible in the case $T_s\gg\eta$, where $T_s$ is the characteristic temperature corresponding to the energy gain in the $\Gamma_4$ triplet state relative to the $\Gamma_1$ level due to the Kondo effect. b) The CEF singlet fixed point is possible in the case $T_s\ll\eta$. There exists the critical value $\eta_c$ that separates a) and b). All the fixed points are explained by the shifts in the f$^2$-levels.

\begin{figure}[t]
	\begin{center}
    \includegraphics[width=.4\textwidth]{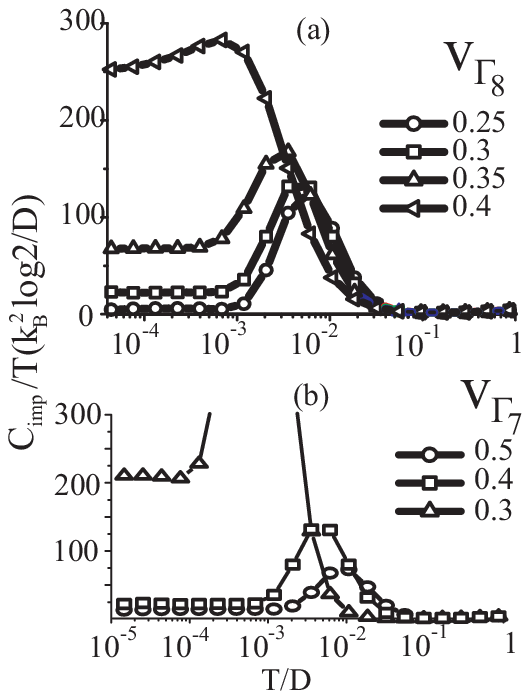}
  \end{center}
\caption{(a) Specific heat divided by temperature, $C_{\rm imp}/T$, vs $T$ for $\Gamma_1$ ground state, for a series of the hybridization $v_{\Gamma_8}$ with other parameters fixed as $v_{\Gamma_7}=0.4$, $\eta=0.0001$, $E_{\Gamma_1}=-1.5$ and $E_{\Gamma_{7(8)}}=-0.9$. $\eta$ is renormalized to $\sim 0.005$ for these parameters. (b) That for a series of the hybridization $v_{\Gamma_7}$, with other parameters fixed as $v_{\Gamma_8}=0.3$, $\eta=0.0001$, $E_{\Gamma_1}=-1.5$ and $E_{\Gamma_{7(8)}}=-0.9$. If we set $D/k_B=10^4$K, $C_{\rm imp}/T\simeq 0.57\times {\rm (value\ of\ ordinate)\ mJ/mol}\cdot{\rm K^2}$ in both (a) and (b).}
\label{fig-gamma}
\end{figure}

 Even in the region of the CEF fixed point, it is interesting to see whether the quasiparticles become heavy or not. We estimate $C_{\rm imp}/T$, specific heat divided by the temperature due to an impurity around the CEF singlet fixed point as shown in Fig. \ref{fig-gamma}. It is a common feature that there exists a Schottky peak in $C_{\rm imp}/T$ that corresponds to releasing $S_{\rm imp}$ as $k_B \log4 \to 0$, and $C_{\rm imp}/T$ approaches a saturated value as $T\to 0$. $C_{\rm imp}/T$ at $T\to 0$ becomes large as $v_{\Gamma_8}$ increases (Fig. \ref{fig-gamma}(a)) and as $v_{\Gamma_7}$ decreases (Fig. \ref{fig-gamma}(b)). These are precisely a reminiscence of the ``S'' fixed point (i.e., the effect of the critical point between the CEF singlet and the NFL ground state in Fig. \ref{fig-PHASE}). It is apparent from Fig. \ref{fig-gamma} that the heaviness of the quasiparticles is due to the hybridization with $\Gamma_8$ conduction electrons. On the other hand, $v_{\Gamma_7}$ works to help in stabilizing the CEF singlet ground state, i.e., it renormalizes $\eta$ to become large. 

\begin{figure}
	\begin{center}
    \includegraphics[width=.43\textwidth]{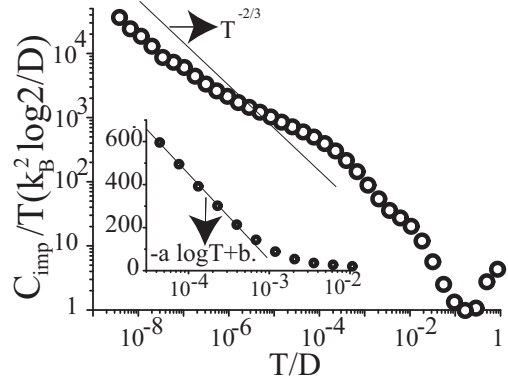}
  \end{center}
\caption{Specific heat divided by temperature, $C_{\rm imp}/T$, for $\Gamma_4$ ground state. The parameters used are $E_{\Gamma_{7(8)}}=-0.2$, $E_{\Gamma_1}=-1.5$, $v_{\Gamma_8}=0.5, v_{\Gamma_7}=0.3$ and $\eta=0.0005$. The result is consistent with the prediction of BCFT, $C_{\rm imp}/T \sim T^{-2/3}$ in the limit $T\to0$. Inset: $C_{\rm imp}/T$ around $T\sim10^{-4}$. It is noted that this region is experimentally attainable and not extremely close to the ``S'' fixed point.}
\label{fig-Sgamma}
\end{figure}

The parameter set of a small $v_{\Gamma_8}$ or a large CEF splitting $\eta$ corresponds to the PrRu$_{4}$Sb$_{12}$ case, and the other limit corresponds to the PrFe$_4$P$_{12}$ case. The heaviness of the quasiparticle of the latter compound is related to the ``S'' fixed point. 
At this ``S'' fixed point, the scaling dimension of the leading irrelevant operator given by BCFT is $1/6$ so that $C_{\rm imp}/T$ shows a divergent behavior as $C_{\rm imp}/T \sim T^{-2/3}$ (see Fig. {\ref{fig-Sgamma}})\cite{KOGA, Hat1}. However, in the experimentally attainable temperature region, we observe the $-\log T$ -like dependence of $C_{\rm imp}/T$, as seen in the inset of Fig. {\ref{fig-Sgamma}}. In this temperature region, our results are similar to the prediction based on the two-channel Kondo model, even though we did not assume that the ground state is the f$^2$-$\Gamma_{23}$ doublet.
It is noted that the origin of the heaviness of the quasiparticles is different from the LS scheme \cite{Otsuki}.

 Finally, let us make a comment about the effect of $T_h$ symmetry. In this paper, we have restricted ourselves to $O_h$ symmetry, because of certain technical reasons.
 The f$^2$-$\Gamma_5$ state with $J=4$ is represented in our pseudospin language as
\begin{eqnarray}
\Gamma_5^+ : \frac{1}{\sqrt{21}}[\frac{1}{2}c_{\downarrow}^{\dagger}a_{\frac{3}{2}}^{\dagger}+\frac{\sqrt{3}}{2}c_{\uparrow}^{\dagger}a_{\frac{1}{2}}^{\dagger}]+\sqrt{\frac{20}{21}}a_{\frac{3}{2}}^{\dagger}a_{-\frac{1}{2}}^{\dagger},\label{G5+}
\end{eqnarray}
\begin{eqnarray}
\Gamma_5^o : \frac{1}{\sqrt{42}}[c_{\downarrow}^{\dagger}a_{-\frac{3}{2}}^{\dagger}-c_{\uparrow}^{\dagger}a_{\frac{3}{2}}^{\dagger}]+\sqrt{\frac{10}{21}}[a_{-\frac{1}{2}}^{\dagger}a_{-\frac{3}{2}}^{\dagger}-a_{\frac{3}{2}}^{\dagger}a_{\frac{1}{2}}^{\dagger}],\label{G50}
\end{eqnarray}
\begin{eqnarray}
\Gamma_5^- : \frac{1}{\sqrt{21}}[\frac{1}{2}c_{\uparrow}^{\dagger}a_{-\frac{3}{2}}^{\dagger}+\frac{\sqrt{3}}{2}c_{\downarrow}^{\dagger}a_{-\frac{1}{2}}^{\dagger}]-\sqrt{\frac{20}{21}}a_{\frac{1}{2}}^{\dagger}a_{-\frac{3}{2}}^{\dagger}.\label{G5-}
\end{eqnarray}
It is noted that states (\ref{G5+})-(\ref{G5-}) are not the eigenstates of $S^{\rm tot}$, $S^{\rm tot}_{z}$ and $h^{\rm tot}$. If we include these states, considering $T_h$ symmetry (under which $\Gamma_4$ and $\Gamma_5$ mix), the ``S'' fixed point should become unstable because anisotropy exists there in the exchange interaction, breaking the SU(2) symmetry. Such features of an unstable NFL fixed point can be seen in a recent NRG study reported in ref. 23, in which they discussed a case including $\Gamma_3$ and $\Gamma_5$ states in addition to the ground state quartet ($\Gamma_1$ and $\Gamma_4$) as the low-energy f$^2$ states (thus breaking the SU(2) symmetry in the present paper). However, the precise identification of the unstable NFL was not performed in the study detailed in ref. 23, although it seems to be related to the ``S'' fixed point discussed in the present paper. Another crucial point is that there is no triplet state constructed by two $\Gamma_7$ electrons even if we work under $T_h$ symmetry. For this reason, the Kondo effect caused by $\Gamma_7$ conduction electrons\cite{Otsuki} is suppressed in the case of the strong spin-orbit interaction we are considering here. From these facts, we can expect that in the case of a strong Hund's rule coupling, it is possible for the Kondo effect of $\Gamma_7$ conduction electrons to occur; on the other hand, in the case of strong spin-orbit interactions, the criticality related to the ``S'' fixed point creates strong renormalizations, if contributions from the $\Gamma_8$ conduction electrons are important.



We would like to thank H. Harima and T. Takimoto for fruitful discussions.
This work is supported by a Grant-in-Aid for Scientific Research 
(No. 16340103), a Grant-in-Aid for Scientific Research in Priority Areas (No. 16037209) and the 21st Century COE Program (G18) of the Japan Society for the 
Promotion of Science.

\end{document}